\documentclass{appolb}
\usepackage{epsfig}
\begin{document}
\title{Public transport systems in Poland: from Bia{\l}ystok to Zielona G\'ora by bus and
tram using universal statistics of complex networks
\thanks{Presented at the 17th Marian Smoluchowski Symposium
on Statistical Physics, Zakopane, Poland, September 4-9 2004}}
\author{Julian Sienkiewicz and Janusz A. Ho{\l}yst
\address{Faculty of Physics and Center of Excellence for Complex
Systems Research, Warsaw University of Technology, Koszykowa 75,
PL-00-662 Warsaw, Poland}}
\maketitle
\begin{abstract}
We have examined a topology of 21  public transport networks in
Poland. Our data exhibit several universal features in considered
systems when they are analyzed from the point of view of evolving
networks. Depending on the assumed definition of a network
topology the degree distribution can follow a power law $p(k) \sim
k^{-\gamma}$  or can be described by an exponential function $p(k)
\sim \exp(-\alpha k)$. In the first case one observes that mean
distances  between two nodes are a linear function of logarithms
of their degrees product.

\end{abstract}

\section{Introduction}
Have you ever been confused and tired using city public transport?
Have you studied city maps during your holiday's excursions
looking for the best connection from a railway station to your
hotel? Do you think  there is any regularity in complex objects
called {\it public transport systems}? During the last few years
several transport networks have already been investigated using
various concepts of statistical physics of  complex networks.  A
study of Boston underground transportation system (MBTA)
\cite{eff, boston} has taken into account physical distances and
has been focused on problems of links  efficiency and  costs. In
\cite{india} a part of the Indian Railway Network (IRN) has been
considered and a new topology describing the system as a set of
train lines, not stops has been introduced. In \cite{vienna} data
from MBTA and U-Bahn network of Vienna have been compared to
predictions of random bipartite graphs. Another class of transport
systems form airport and airlines networks: world-wide
\cite{airport, architect}, Chinese \cite{china_air} or Indian
\cite{india_air} airline networks - here such properties as links
weights, correlations among different airports and their
centrality have been investigated.

 In the present paper we have studied a part of data for
public transport networks in $21$ Polish cities and we find that
among apparent differences there are also universal features of
these systems.  As far as we know, our results are the first
comparative survey of several individual transport systems in the
same country.

\section{Network topology: space L and space P}\label{sec:deflp}

It is clear that distances for city travelers are not the same as
physical distances if he/she needs to come from a city point $A$
to a city point $B$  using existing public transport media.
Sometimes it occurs that a physical distance between points $A$
and $B$ is not very large but  the travel between these  points in
the city is unfortunately  time consuming since either a direct
bus makes a lot of loops on its way or we need to change buses or
tramways several times. It follows that one can introduce at least
two different representations of city transport networks where a
network is a set of nodes (vertices) and links (edges) between
them.
\begin{figure}[h]
 \centerline{\epsfig{file=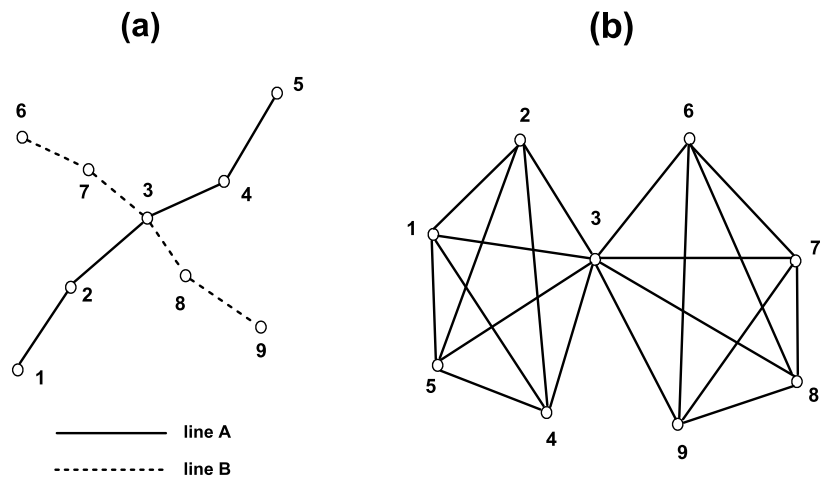,width=.9\columnwidth}}
    \caption{Transformation from space L (a) to space P (b) using an example of
     two public transport lines A and B.}
    \label{fig:przestrzenie}
\end{figure}
The first representation is the space L which consists of nodes
being bus or tramway stops while a link between two nodes exists
provided they are consecutive stops on a bus or a tramway line.
The distance in such a space is measured by the total number of
stops passed on the shortest path between two nodes. However the
distance measured in such a way does not reflect the necessity of
transfer during the trip. This factor is taken into account in the
second space P \cite{india}. Nodes in such a space are the same as
in the previous one but now an edge connecting two nodes means
that there is a link by a {\it single} bus or tramway between
them. It follows that in the space P the distances are numbers of
{\it transfers} (plus one) needed during the trip. It is obvious
that distances defined in the space P are much shorter than in the
space L and there is no universal relation between them. Both
spaces are presented at Fig. \ref{fig:przestrzenie}.

\section{Explored systems}

\begin{figure}[ht]
 \centerline{\epsfig{file=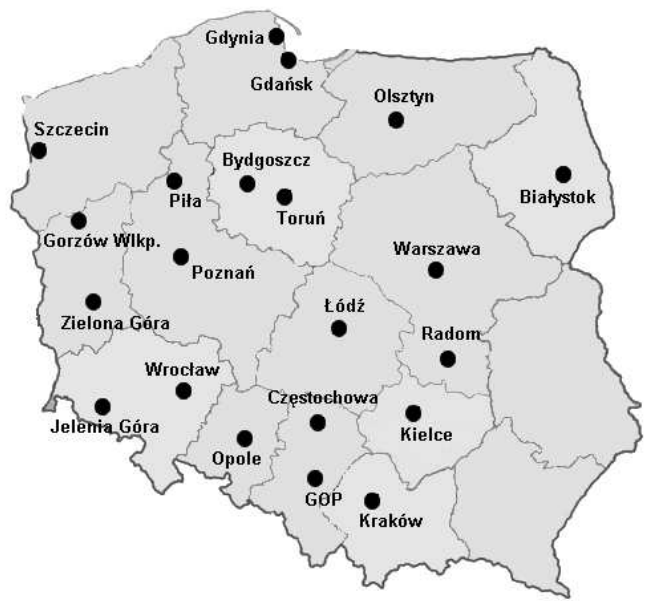,width=.7\columnwidth}}
    \caption{Map of twenty one examined cities in Poland.}
    \label{fig:mapa}
\end{figure}

We have studied data collected from 21 public transport networks
in Polish cities that are depicted at Fig. \ref{fig:mapa} and
listed in the Table \ref{tabela}. The first analyzed features are
degree distributions for networks represented in both spaces. A
degree of node $i$ is the number $k_i$ of its nearest neighbors.
In regular networks (e.g. crystals) all nodes (e.g. atoms) can
have the same degree. In complex networks \cite{bap} there is a
wide spectrum of degrees and a large interest on such systems
started from a discovery of scale-free distributions in several
real networks \cite{watts,badiam,ba,new2}.

\begin{figure}[!ht]
 \centerline{\epsfig{file=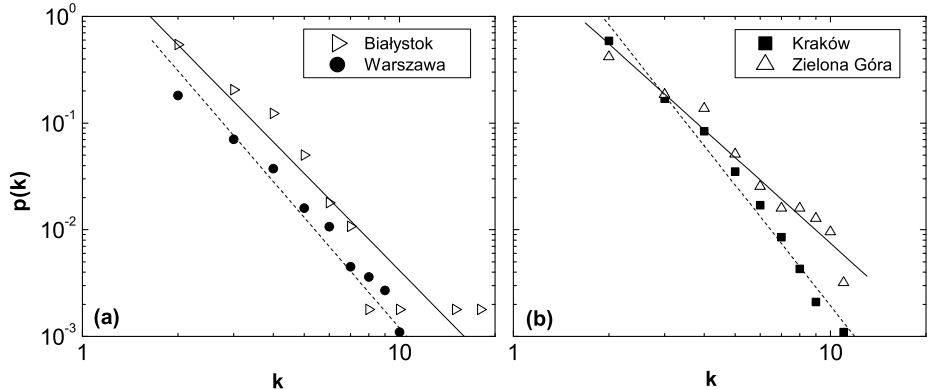,width=\columnwidth}}
    \caption{Degree distribution in space L with power law $k^{-\gamma}$ fit. (a) Bia{\l}ystok $\gamma=3.0 \pm
    0.4$ (solid line) and Warszawa $\gamma=3.44 \pm 0.22$ (dotted line). (b) Krak\'ow $\gamma=3.77 \pm 0.18$
    (solid line) and Zielona G\'ora $\gamma=2.68 \pm 0.20$ (dotted line).}
    \label{fig:pl}
\end{figure}

The Fig. \ref{fig:pl} shows typical plots for degree distribution
in the space L. In all studied plots we neglected the point $k=1$
that corresponds to line ends. Remaining parts of degree
distributions can be approximately described by a power law

\begin{equation}
p(k)\sim k^{-\gamma}
\label{gamma}
\end{equation}

with a characteristic exponent $\gamma$ between $2.4$ and $4.1$.
Values of exponents $\gamma$ are different from the value
$\gamma=3$ which is characteristic for Barab\'asi-Albert model of
evolving networks with preferential attachment and one can suppose
that a corresponding model for transport network evolution should
include several other effects. One can see also that  larger
exponents $\gamma$ correspond usually to larger numbers $N$ of
nodes in the network (Table \ref{tabela}).

\begin{figure}[ht]
 \centerline{\epsfig{file=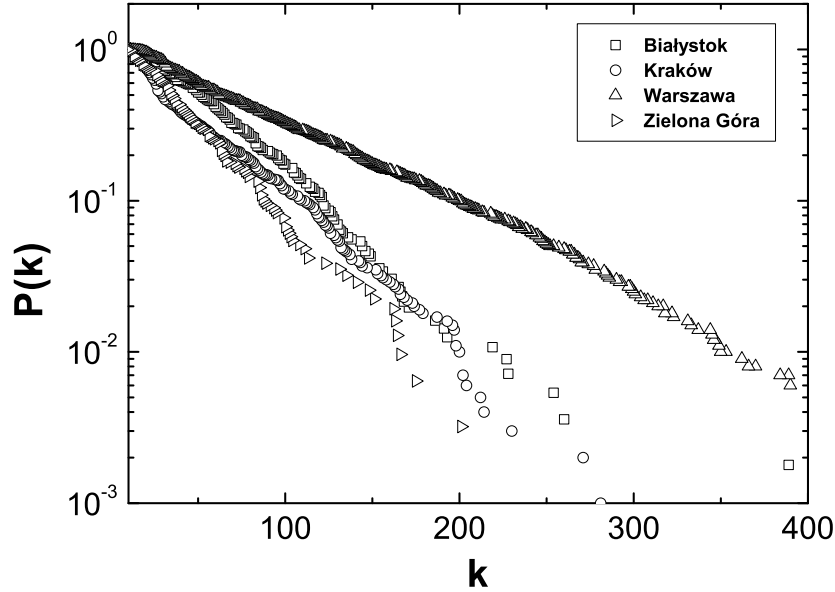,width=.9\columnwidth}}
    \caption{Cumulative distribution $P(k)$ in space P for Bia{\l}ystok, Krak\'ow, Warszawa and Zielona G\'ora.}
    \label{fig:pk}
\end{figure}

A quite other situation is in the space P. Corresponding
cumulative degree distributions for selected cities
$P(k)=\int^{k_{max}}_k p(k')dk'$ are presented at Fig.
\ref{fig:pk}. The distributions $P(k)$ and $p(k)$ are well fitted
by exponential representation

\begin{equation}
p(k) \sim \exp(-\alpha k).
\label{alpha}
\end{equation}

As it is well known \cite{bamf} the exponential distribution
(\ref{alpha}) can occur when a network evolves but nodes are
attached completely {\it randomly}. We are surprised that such a
random evolution could well  correspond to the topology  of urban
transport networks.

Table \ref{tabela} presents  exponents  $\gamma$ and $\alpha$ for
investigated cities. The values have been received from the
standard linear regression method.
\begin{table}[!ht]
\setlength{\tabcolsep}{6pt}
  \centering
  \begin{tabular}{ccccccccc}
    \hline \hline
    city & $N$ & $\gamma$ & $\Delta \gamma$ & $R_{\gamma}^2$ & $\alpha$ & $\Delta \alpha$ & $R_{\alpha}^2$ & $R_{l}^2$\\
    \hline
Bia{\l}ystok & 559 & 3.0 & 0.4 & 0.945 & 0.0211 & 0.0002 & 0.997 & 0.873\\
Bydgoszcz & 276 & 2.8 & 0.3 & 0.961 & 0.0384 & 0.0004 & 0.996 & 0.965\\
Cz\c{e}stochowa & 419 & 4.1 & 0.4 & 0.974 & 0.0264 & 0.0004 & 0.992 & 0.976\\
Gda\'nsk & 493 & 3.0 & 0.3 & 0.952 & 0.0304 & 0.0006 & 0.981 & 0.980\\
Gdynia & 406 & 3.04 & 0.2 & 0.983 & 0.0207 & 0.0003 & 0.990 & 0.967\\
GOP & 2811 & 3.46 & 0.15 & 0.987 & 0.0177 & 0.0002 & 0.988 & 0.885\\
Gorz\'ow Wlkp. & 269 & 3.6 & 0.3 & 0.983 & 0.0499 & 0.0009 & 0.994 & 0.984\\
Jelenia G\'ora & 194 & 3.0 & 0.3 & 0.979 & 0.038 & 0.001 & 0.984 & 0.994\\
Kielce & 414 & 3.00 & 0.15 & 0.992 & 0.0263 & 0.0004 & 0.991 & 0.963\\
Krak\'ow & 940 & 3.77 & 0.18 & 0.992 & 0.0202 & 0.0002 & 0.996 & 0.977\\
{\L}\'od\'z & 1023 & 3.9 & 0.3 & 0.968 & 0.0251 & 0.0001 & 0.998 & 0.983\\
Olsztyn & 268 & 2.95 & 0.21 & 0.980 & 0.0226 & 0.0004 & 0.986 & 0.985\\
Opole & 205 & 2.29 & 0.23 & 0.978 & 0.0244 & 0.0004 & 0.989 & 0.992\\
Pi{\l}a & 152 & 2.86 & 0.17 & 0.990 & 0.0310 & 0.0006 & 0.989 & 0.989\\
Pozna\'n & 532 & 3.6 & 0.3 & 0.978 & 0.0276 & 0.0003 & 0.994 & 0.976\\
Radom & 282 & 3.1 & 0.3 & 0.960 & 0.0219 & 0.0004 & 0.989 & 0.991\\
Szczecin & 467 & 2.7 & 0.3 & 0.963 & 0.0459 & 0.0006 & 0.995 & 0.979\\
Toru\'n & 243 & 3.1 & 0.4 & 0.964 & 0.0331 & 0.0006 & 0.990 & 0.979\\
Warszawa & 1530 & 3.44 & 0.22 & 0.980 & 0.0127 & 0.0001 & 0.998 & 0.985\\
Wroc{\l}aw & 526 & 3.1 & 0.4 & 0.964 & 0.0225 & 0.0002 & 0.993 & 0.983\\
Zielona G\'ora & 312 & 2.68 & 0.20 & 0.979 & 0.0286 & 0.0003 & 0.995 & 0.996\\
\hline \hline
\end{tabular}
\caption{Number  of nodes $N$, coefficients  $\gamma$ and $\alpha$
with their standard errors $\Delta \gamma$, $\Delta \alpha$ and
Pearson's coefficients $R_{\gamma}^2$, $R_{\alpha}^2$, for
considered cities. The last column $R_{l}^2$ represents Pearson's
coefficient for the scaling (\ref{AB}). Fitting to the scaling
relations (\ref{gamma}) and (\ref{AB}) has been performed at whole
ranges of degrees $k$. Fitting to (\ref{alpha}) has been performed
at approximately half of available ranges to exclude large
fluctuations occurring for higher degrees (See Fig.
\ref{fig:pk}).} \label{tabela}
\end{table}

\subsection{Path length as function of product $k_ik_j$}\label{sec:l}

In \cite{agatac} an analytical model of average path length was
considered and  it was shown  that the shortest path length
between nodes $i$ and $j$ possessing degrees $k_i$ and $k_j$  in a
random graph characterized by its degree distribution $p(k)$ can
be described as:

\begin{equation}\label{eq:lij}
    l_{ij}(k_i,k_j) = \frac{-\ln k_ik_j+\ln
    \left(\langle k^2 \rangle - \langle k \rangle \right)+\ln N - \gamma}{\ln
    \left(\langle k^2 \rangle / \langle k \rangle-1 \right)}+\frac{1}{2}
\end{equation}

where $\gamma = 0.5772$ is Euler constant while $\langle k
\rangle$ and $\langle k^2 \rangle$ are corresponding first and
second moments of $p(k)$ distributions. In
\cite{nasz_physa,nasz_prl,moter} a random tree (a random graph
with no loops) was studied  and it was shown that

\begin{equation}\
    l_{ij}(k_i,k_j) = A -B \log k_ik_j   \label{AB}
\end{equation}

where coefficients $A$ an $B$ depend on an average branching
factor $\kappa$ of the considered tree and on a  total number of
its edges $E=N \langle k \rangle/2$

\begin{equation}
A=1+ \frac{\log(2 E)}{\log \kappa}
\end{equation}
\begin{equation}
B=\frac{1}{\log \kappa}
\end{equation}

\begin{figure}[!ht]
 \centerline{\epsfig{file=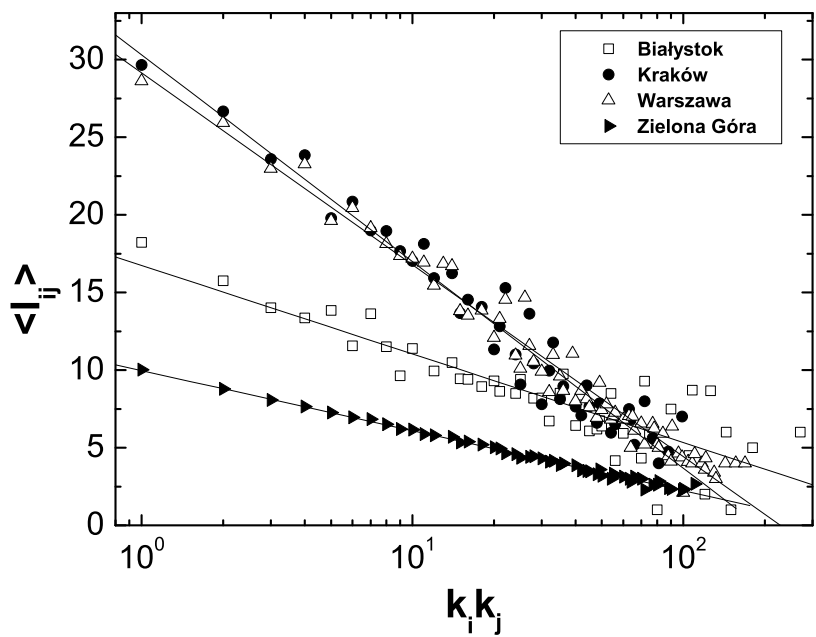,width=.9\columnwidth}}
     \caption{Dependence of $l_{ij}$ on $k_ik_j$ in space L for Bia{\l}ystok, Krak\'ow, Warszawa and Zielona G\'ora.}
    \label{fig:lij}
\end{figure}

We have found that if  distances are measured in the space L then
the scaling (\ref{AB}) is well fulfilled for considered public
transport networks (Fig. \ref{fig:lij}). Table \ref{tabela}
presents corresponding Pearson's coefficients. One can see that
except the cases of Bia{\l}ystok and GOP all other $R_l^2$
coefficients are above $0.96$ and the best fit to the scaling
relation (\ref{AB}) has been found for Zielona G{\'ora}, where
$R^2=0.996$. Observed values of $A$ and $B$ coefficients differ as
much as 20 percent (in average) from theoretical values received
for random graphs where contribution from clustering and node
degree correlations are taken into account  (see \cite{nasz_physa,
nasz_prl, nasz_pre}).

It is useless to examine the relation (\ref{AB})  in the space P
because of the structure of this space. In fact the set $l_{ij}$
contains usually only 3 points what means that one needs just two
changes of a bus or a tram to come from one city point to another
\cite{nasz_pre}.

\section{Conclusions}
In conclusion we  have observed that public transport networks in
many Polish cities follow  universal scalings. The degree
distribution $p(k)$ fits to  a power law in the space L where a
distance is measured in numbers of bus or tram stops. A
corresponding distribution in the space P where a distance is
measured in a number of transfers between different vehicles
follows an exponential distribution. Distances in the space L are
a linear function of logarithms of corresponding nodes degrees.

\section{Acknowledgments}
JAH is thankful to Professor Andrzej Fuli{\'n}ski for many useful
discussions   and for a creative atmosphere established during all
Marian Smoluchowski Symposia on Statistical Physics in Zakopane.

\end{document}